\documentclass[USenglish]{article}	
\usepackage[utf8]{inputenc}				%(only for the pdftex engine)
\usepackage{lmodern} 
\usepackage{microtype}
\usepackage[numbers,square,sort&compress]{natbib}
\usepackage{caption}
\usepackage{subcaption}

\usepackage{amsmath}
\usepackage[utf8]{inputenc}
\DeclareMathOperator\erf{erf}
\DeclareMathOperator\Var{Var}
\DeclareMathOperator\Cov{Cov}
\DeclareMathOperator\E{E}
\usepackage[toc,title,page]{appendix}
\usepackage{multirow}

\usepackage{booktabs}
\usepackage{hyperref}
\usepackage[margin=1in]{geometry}
\usepackage{graphicx}
\begin{document}

\title{Static quantification of player value for fantasy basketball}
%\subtitle{Insert subtitle if needed}

\author{Zach Rosenof}

\maketitle

\includegraphics[scale = 0.5]{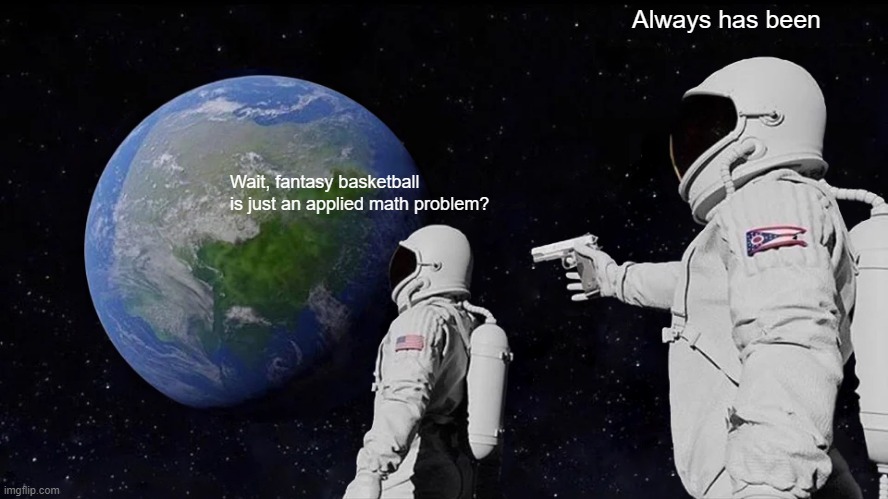}

\abstract{Fantasy basketball has a rich underlying mathematical structure which makes optimal drafting strategy unclear. A central issue for category leagues is how to aggregate a player's statistics from all categories into a single number representing general value. It is shown that under a simplified model of fantasy basketball, a novel metric dubbed the ``G-score'' is appropriate for this purpose. The traditional metric used by analysts, ``Z-score'', is a special case of the G-score under the condition that future player performances are known exactly. The distinction between Z-score and G-score is particularly meaningful for head-to-head formats, because there is a large degree of uncertainty in player performance from one week to another. Simulated fantasy basketball seasons with head-to-head scoring provide evidence that G-scores do in fact outperform Z-scores in that context.}
	
\section{Introduction}

Fantasy basketball is a game in which participants manage virtual basketball teams linked to the performances of real players. According to numbers from the Fantasy Sports \& Gaming Association, approximately 20 million Americans and Canadians played fantasy basketball in 2022 (FSGA, 2023). 

Most fantasy basketball leagues have category-based scoring (Bartilotta, 2023). Despite the broad popularity of fantasy basketball, the literature on so-called category leagues is scant. Quantitatively, the fantasy basketball community relies on a heuristic metric called Z-score to estimate the values of different players for them (Kelly, 2021). As will be shown, the metric can be justified under certain assumptions, but is missing a crucial element

\subsection{Player selection}

In fantasy basketball, ``managers'' draft actual players to their ``teams'' before the NBA or WNBA season. Like in a real sports draft, no player can be selected more than once, and once a player gets picked by a manager they are removed from the draft board. Usually, the picks are made in a snaking order as depicted in Table \ref{fig:Snake}. Alternatively, players can be chosen via a mock auction for which all managers start with the same amount of virtual money (ESPN, 2023). 

Teams all have the same number of players, often thirteen. During the season, managers can swap their players with ``free agents'' or trade them to other managers. The exact rules for how managers are allowed to change their team during the season differ from league to league

\begin{table}[!ht]
\centering
\begin{tabular}{r r r r r r}
Team A & Team B & Team C & Team D & Team E & Team F \\ \midrule
1 & 2 & 3 & 4 & 5 & 6 \\
12 & 11 & 10 & 9 & 8 & 7 \\
13 & 14 & 15 & 16 & 17 & 18 \\
24 & 23 & 22 & 21 & 20 & 19 \\
\end{tabular}
\caption{Draft order for a six-team, four-round snake draft}
\label{fig:Snake}
\end{table} 

\subsection{Scoring}

Most fantasy basketball leagues are category leagues, meaning that they score managers based on how many categories they win against each other (Bartilotta, 2023). The three main formats for category leagues are

\begin{itemize}

\item Rotisserie: Teams are ranked according to aggregate performance in each statistical category over the course of a season. The team with the best average ranking across categories wins 

\item Head-to-Head: Each Category: Teams are matched up against opponents each week and their category-level results are tabulated. E.g. if a team wins six categories, loses two, and ties one, they score 6-2-1. Teams are ranked at the end of the season by aggregate records

\item Head-to-Head: Most Categories: Teams are matched up against opponents each week and the team that wins a majority of categories wins the week. Teams are ranked at the end of the season by number of weeks won (ESPN, 2023)

\end{itemize}

Common settings specify nine categories: points, rebounds, assists, steals, blocks, three-pointers, field goal \%, free throw \%, and turnovers. Player performances are aggregated to get overall team metrics for each scoring period. Points, rebounds, assists, steals, blocks, three-pointers, and turnovers, henceforth dubbed ``counting statistics'' are aggregated by taking the sum. Field goal \% and free throw \%, henceforth dubbed ``percentage statistics'' are aggregated by taking the total number of successes over the total number of attempts for the category. For Rotisserie, player performances are aggregated over the course of an entire season. For head-to-head, scoring periods are usually weeks. Categories are won by the team with the higher metric, except for the turnovers category which is won by the team with the lower metric. Examples of how the scoring system works are included as Tables \ref{fig:Aggregation}, \ref{fig:Roto}, \ref{fig:Each Category}, and \ref{fig:Most Categories}

\begin{table}[!ht]
\centering
\begin{tabular}{l l r r r}
 & & Points & Turnovers & Free throw \% \\ \midrule
 Team A & Player 1 & 10 & 1 & 66\% (2/3) \\
& Player 2 & 15 & 2 & 100\% (4/4) \\
& Player 3 & 33 & 2 & 50\% (5/10) \\
\cline{2-5}
& Aggregate & \textbf{58} & 5 & 64\% (11/17) \\
& & & & \\
Team B & Player 4 & 12 & 1 & 50\% (1/2) \\
& Player 5 & 8 & 1 & 0\% (0/1) \\
& Player 6 & 20 & 1 & 80\% (8/10) \\
\cline{2-5}
& Aggregate & 40 & \textbf{3} & \textbf{69\% (9/14)}  \\
\end{tabular}
\caption{Example of how metrics are aggregated and compared. Team A wins points and loses the two other categories. It loses turnovers because for that category, the team with the lower total wins}
\label{fig:Aggregation}
\end{table}

\begin{table}[!ht]
\centering
\begin{tabular}{l r r r r r}
& Points rank & Blocks rank & Rebounds rank & Total & Result \\ \midrule
Team A & 1 & 3 & 3 & 7 & Second\\
Team B & 2 & 2 & 1 & 5 & First\\
Team C & 3 & 1 & 4 & 8 & Third\\
Team D & 4 & 4 & 2 & 10 & Fourth\\
\end{tabular}
\caption{Example of rotisserie scoring with four teams and three categories. Ranks are calculated by totals across a full season- 1 is best, 4 is worst}
\label{fig:Roto}
\end{table}

\begin{table}[!ht]
\begin{subtable}[h]{1\linewidth}
\centering
\begin{tabular}{l l l l l l}
& Week 1 & Week 2 & Week 3 & Total & Result \\ \midrule
Team A & 5-4 & 1-8 & 4-5 & 10-17 & Fourth\\
Team B & 4-5 & 4-5 & 3-6 & 11-16 & Third\\
Team C & 3-6 & 8-1 & 6-3 & 17-10 & First\\
Team D & 6-3 & 5-4 & 5-4 & 16-11 & Second\\
\end{tabular}
\caption{Each Category scoring}
\label{fig:Each Category}
\end{subtable}

\begin{subtable}[h]{1\linewidth}
\centering
\begin{tabular}{l l l l l l}
& Week 1 & Week 2 & Week 3 & Total & Result \\ \midrule
Team A & 5-4 (W) & 1-8 (L) & 4-5 (L) & 1-2 & Third\\
Team B & 4-5 (L) & 4-5 (L) & 3-6 (L) & 0-3 & Fourth\\
Team C & 3-6 (L) & 8-1 (W) & 6-3 (W) & 2-1 & Second\\
Team D & 6-3 (W) & 5-4 (W) & 5-4 (W) & 3-0 & First\\
\end{tabular}
\caption{Most Category scoring}
\label{fig:Most Categories}
\end{subtable}

\caption{Scoring with four teams, nine categories, and three weeks}
\end{table}

\subsection{Position rules}

On a daily basis, players must be placed at positions in order for their games to count. In fantasy basketball the main positions are center, point guard, shooting guard, power forward, and small forward. Each player has a set of positions for which they are eligible. Teams are limited to a certain number of players at each position for day, and when no slots are available, players must be placed ``on the bench'' where their games do not count

\section{Z-scores for static value}

A core component of fantasy basketball strategy is having an informed opinion of which players are best to draft. 

To this end, many fantasy basketball experts e.g. Basketball Monster publish estimates of player value before each season, based on how they expect players to perform (RotoMonster, no date). These value estimates do not depend on draft circumstances, which make them convenient and easy to use. Managers can apply them during snake drafts by taking players with the highest available values, without needing to adjust for previously drafted players. Managers can also use them for auctions, converting them to dollar values by subtracting out replacement value (the value of a player that can be added for free after the auction) then re-scaling to the amount of money available (Roberts, 2014). The resulting dollar values can be used as benchmarks for how much they should be willing to spend on players. 

Naturally, experts who make value estimates often try to do so in a mathematical way. This is non-trivial for category leagues, because there is no obvious way to aggregate expected performances across categories. 

Z-scoring is the traditional method for static value quantification used by Basketball Monster and others (Kelly, 2021). It is a three-step process

\begin{enumerate}
\item For percentage statistics, multiply expected performances by expected volume, divided by average volume
\item For all categories, subtract out the mean and divide by the standard deviation
\item Sum up all of the category-level scores to get scores for each player 
\end{enumerate}

This approach is attractive because it ensures that the distributions per category all have a mean of zero and standard deviation of one, which intuitively seems reasonable for a scoring system. However, being intuitively reasonable is not proof that Z-scores are optimal. Understanding the sense in which Z-scores are or are not optimal requires mathematical analysis

\section{Justifying G-scores}

Under a synthetic version of the static value problem which is described in this section,, Z-scores are close to optimal but not fully general. A novel metric dubbed G-score fixes Z-score by extending it to incorporate uncertainty in player performances 

\subsection{Model} \label{Model}

Consider a manager whose turn it is to select a player. Assume that
\begin{itemize}

\item Each managers' objective $V$ is to have the highest possible expected value of number of categories won against an arbitrary opponent, across an arbitrary scoring period. The set of categories is $C$, of size $|C|$
\item All players on all teams, except for the one being currently selected, are chosen randomly from set $Q$. $Q$ is composed of $|Q|$ players, where $|Q|$ is the number of players that will be drafted into the league. E.g. if twelve managers are each picking thirteen players, $|Q| = 12 * 13 = 156$
\item Performances for all players are chosen randomly from a set of possible outcomes $W$ of size $|W|$. The selection process is independent for each player
\item All games played by all chosen players, $N$ per team, count towards team totals
\item A player's value is proportional to $V_p$ - $V_0$, where $V_p$ is the objective value given that player $p$ is chosen, and $V_0 = \frac{|C|}{2}$, the default objective value 
\end{itemize}

The question is how player value should be calculated under these conditions. How well the conditions represent real fantasy basketball is discussed in Section \ref{Ass}

\subsection{Solution}

It is posited that G-scores, as defined by Figure \ref{fig:GScore}, are optimal under the conditions of the simplified model of fantasy basketball laid out in Section \ref{Model}. Essentially, G-scores are the same as Z-scores except the denominators are modified to add period-to-period variance to the player-to-player-variance already included in Z-scores. Z-scores are equivalent to G-scores with $|W| = 1$, removing all period-to-period variance

\begin{table}[!ht]
\begin{subtable}[h]{1\linewidth}
\begin{tabular}{| l | l |}
\hline
& \\[-5pt]
$\theta_M \left( q,w \right)$ & Performance for player $q$ in $w$ \\[5pt]
& \\[-5pt]
$\mu_M\left( q \right)$ & Mean of $\theta_M \left( q,w \right)$ for player $q$ across $W$: $\frac{\sum_{w \in W} \theta_M \left( q,w \right)}{|W|} $\\[5pt]
& \\[-5pt]
$\mu_M$ & Mean of $\mu_M\left( q \right)$ across players: $\frac{\sum_{q \in Q} \mu_M\left( q \right)}{|Q|} $\\[5pt]
$\sigma_M$ & Standard deviation of $\mu_M\left( q \right)$ across $Q$: $\sqrt{\frac{\sum_{q \in Q}  \left( \mu_M\left( q \right) - \mu_M \right) ^2}{|Q|}}$ \\[5pt]
$\tau_M \left( q \right)$ & Standard deviation of $\theta_M \left( q,w \right)$ across $W$: $\sqrt{\frac{\sum_{w \in W}  \left( \theta_M \left( q,w \right) - \mu_M\left( q \right) \right) ^2}{|W|}}$ \\
& \\[-5pt]
$\tau_M$ & Root-mean-square of $\tau_M \left( q \right)$: $\sqrt{\frac{\sum_{q \in Q} \tau_M \left( q \right)^2}{|Q|}} $ \\[10pt]
$\kappa$ & Kappa factor: $\frac{2N}{2N-1}$ \\[5pt]
\hline
\end{tabular}

\caption{Definitions for counting statistics, used for G-scores}
\label{fig:GCountDef}
\end{subtable}

\begin{subtable}[h]{1\linewidth}
\begin{tabular}{| l | l |}
\hline
& \\[-5pt]
$\theta_A \left( q, w \right) $ & Number of attempts made by player $q$ in $w$\\ [5pt]
& \\[-5pt]
$\theta_R \left( q, w \right) $ & Success rate for player $q$ in $w$\\[5pt]
& \\[-5pt]
$\mu_A\left( q \right)$ & Mean of $\theta_A \left( q,w \right)$ for player $q$ across $W$: $\frac{\sum_{w \in W} \theta_A \left( q,w \right)}{|W|} $\\[5pt]
& \\[-5pt]
$\mu_A$ & Mean of $\mu_A\left( q \right)$ across players: $\frac{\sum_{q \in Q} \mu_A\left( q \right)}{|Q|} $\\[5pt]
& \\[-5pt]
$\mu_R\left( q \right) $ & Composite success rate for player $q$ across $W$: $\frac{\sum_{w \in W} \theta_A \left( q, w \right)  \theta_R \left( q, w \right) }{\sum_{w \in W} \theta_A \left( q, w \right) }$\\[5pt]
& \\[-5pt]
$\mu_R$ & Composite success rate across players: $\frac{\sum_{q \in Q}  a_q  \mu_R\left( q \right)}{ \sum_{q \in Q} a_q }$\\[5pt]
& \\[-5pt]
$\sigma_R$ & Standard deviation of $\mu_R\left( q \right)$ across $Q$: $\sqrt{\frac{\sum_{q \in Q}  \left( \mu_R\left( q \right) - \mu_R \right) ^2}{|Q|}}$ \\[10pt]
$\tau_R \left( q \right)$ & Standard deviation of $\frac{\theta_A \left(q,w \right) }{\mu_A} *  \left( \theta_R \left( q, w \right)  - \mu_R \right) $ across $W$ for player q\\[5pt]
& \\[-5pt]
$\tau_R$ & Root-mean-square of $\tau_R \left( q \right)$: $\sqrt{\frac{\sum_{q \in Q} \tau_R \left( q \right)^2}{|Q|}} $  \\[10pt]
\hline
\end{tabular}
\caption{Definitions for percentage statistics, used for G-scores}
\label{fig:GPercDef}
\end{subtable}

\renewcommand{\arraystretch}{2}

\begin{subtable}[!ht]{1\linewidth}
\begin{tabular}{| c | c | c | c |}
\hline
        Points/Rebounds/Assists/Steals/Blocks/Threes & Turnovers  & FG \% / FT \%	\\ 
        \hline
         $\frac{ \mu_M\left( q \right) - \mu_M}{\sqrt{ \left( \sigma_M^2 + \kappa * \tau_M^2 \right) } }$ & $\frac{\mu_M - \mu_M\left( q \right)}{\sqrt{ \left( \sigma_M^2 + \kappa * \tau_M^2 \right) }}$ & $\frac{ \frac{a_q}{\mu_A} *  \left( \mu_R\left( q \right) - \mu_R \right) }{ \sqrt{r_{\sigma}^2 + \kappa * \tau_R^2}}$  \\[5pt]
\hline
\end{tabular}
\caption{Definition of G-score per category. The aggregate G-score is the sum across categories}
\label{fig:GScore}
\end{subtable}

\caption{Definitions for G-scoring}

\end{table}

\subsection{Proof}

A detailed proof for why the G-score solves the problem of Section \ref{Model} is included in the appendix. More briefly, all players contribute period-to-period variance $\tau_M^2$, and all players except the player in question contribute player-to-player variance $\sigma_M^2$. Invoking the central limit theorem, the point differential between two teams for a counting statistic can then be written as

$$
\mathcal{N} \left( \mu_M - \mu_M\left( p \right),  \left( 2N - 1 \right)  * \sigma_M^2 + \left( 2N \right) * \tau_M^2 \right)  
$$

If this value is below zero, than team $A$ scores more points than team $B$ and therefore wins the category. The probability that this value is below zero is its cumulative distribution function at zero. Invoking the Taylor Series approximation for the CDF of the normal distribution and some algebraic manipulation, that probability is

$$
   \frac{1}{2} \left[ 1 +\frac{1}{\sqrt{\pi * \left( N - \frac{1}{2} \right) }} *\frac{ \mu_M\left( p \right) - \mu_M}{ \sqrt{\sigma_M^2 + \frac{2N}{2N-1} * \tau_M^2 }} \right]
$$

This includes the formula for G-score for counting statistics. Applying the same procedure to the percentage statistics is more complicated, but results in the equivalent

$$
   \frac{1}{2} \left[ 1 + \frac{1}{\sqrt{\pi \left(N - \frac{1}{2} \right) }} * \frac{ \frac{\mu_A\left( p \right)}{\mu_A}  \left( \mu_R\left( p \right) - \mu_R \right) }{\sqrt{  \sigma_R^2 + \frac{2N}{2N-1} \tau_R^2 }} \right]
$$

Which includes the G-score for percentage statistics. The manager's objective function $V$ is the sum of winning probabilities across categories, so it is 

$$
   V = \frac{1}{2} \left[|C| + \frac{1}{\sqrt{\pi * \left( N - \frac{1}{2} \right)}} * \sum_{c \in C} G_{pc} \right ]
$$

Where $G_{pc}$ is player $p$'s G-score for category $c$. Subtracting out the default expected number of wins yields 

$$
   V - V' = \frac{1}{2 \sqrt{\pi * \left( N - \frac{1}{2} \right)}} * \sum_{c \in C} G_{pc}
$$

This is player value, as defined in Section \ref{Model}.  G-score is clearly proportional to it, so G-score meets the criteria for player value as well

\subsection{Simulation} \label{Sim}

Simulations were run to test the efficacy of G-scoring. The simulated seasons were twelve-team, thirteen-player head-to-head competitions. Each player's performance in each simulated season was generated by randomly sampling twenty weeks from their actual performances, excluding weeks for which they were injured. Players were only included if they had at least 10 non-injured weeks. Teams were paired against each other and winners were decided by which team had the most points by the end of the twenty week season. All games played by all players counted. Managers had access to weekly performance numbers for each player, allowing them to calculate all relevant metrics for all players. Z-scores based on the full league were used to choose $Q$, based on which managers calculated Z-scores and G-scores. The $\kappa$ factor was approximated at $1$. For the Z vs G matchup, Z-scoring was was tried at each ``seat'' (position in the draft order) of a snake draft, against a field of drafters using G-score. One thousand of the twenty-week simulated seasons were run for every draft seat, allowing for robust estimates of how well that strategy really would have performed in that situation with error bars no greater than 

$$
\frac{\sqrt{1000 * \frac{1}{2}{(1 - \frac{1}{2})}}}{1000} \approx 1.6\%
$$

The same was also done in reverse, with G-score as the single drafter against a field of Z-score drafters.

The results are shown in Table \ref{ZvG}. It demonstrates that drafters selecting players via Z-score consistently performed poorly against opponents using G-score, always scoring far below the baseline expected rate of $\frac{1}{12}$ or 8.33\%. In the reverse scenario, G-score performed much better. It won $21.4\%$ of simulated Each Category seasons, and $32.5\%$ of Most Categories seasons. 

\begin{table}[!ht]
\centering
    \begin{tabular}{r r r r r}
     & \multicolumn{2}{l}{Z vs. G} & \multicolumn{2}{l}{G vs. Z} \\
    & Most Categories & Each Category & Most Categories & Each Category 	\\ \midrule
    Aggregate & 0.4\% & $0.5\%$ & $32.5\%$ & $21.4\%$ \\
    Seat 0 & $0.5\%$ & $0.8\%$ & $41.6\%$ & $30.7\%$ \\
    Seat 1 & $0.2\%$ & $0.3\%$ & $21.1\%$ & $14.5\%$ \\
    Seat 2 & $0.3\%$ & $0.8\%$ & $33.7\%$ & $29.2\%$ \\
    Seat 3 & $0.5\%$ & $0.7\%$ & $30.7\%$ & $19.1\%$ \\
    Seat 4 & $0.3\%$ & $0.4\%$ & $22.7\%$ & $14.1\%$ \\
    Seat 5 & $0.0\%$ & $0.1\%$ & $25.6\%$ & $14.5\%$ \\
    Seat 6 & $0.1\%$ & $0.4\%$ & $30.6\%$ & $17.2\%$ \\
    Seat 7 & $0.4\%$ & $0.1\%$ & $53.0\%$ & $33.3\%$ \\
    Seat 8 & $0.2\%$ & $0.0\%$ & $48.8\%$ & $38.6\%$ \\
    Seat 9 & $0.5\%$ & $0.0\%$ & $33.4\%$ & $16.9\%$ \\
    Seat 10 & $1.1\%$ & $0.8\%$ & $22.2\%$ & $11.4\%$ \\
    Seat 11 & $0.8\%$ & $1.4\%$ & $26.8\%$ & $17.1\%$ \\
     \end{tabular}
    \caption{Win rates of Z-scores and G-scores against each other}
    \label{ZvG}
\end{table}

\section{Discussion}

\subsection{Model assumptions} \label{Ass}

Of course, the model presented in Section \ref{Model} to justify G-scores is not a perfect replica of how fantasy basketball works in practice. The degree to which the model represents true fantasy basketball merits discussion

\subsubsection{Players chosen randomly}

While no manager would pick players truly at random, if all managers picked players purely based on a single ranking list with no consideration for team fit, it stands to reason that their per-category totals would be close to random. A manager's category totals would result from the characteristics of the players that happened to be at the top of the ranking list for their draft slots. 

The simulation results provide evidence that the assumption of players being chosen randomly was not problematic for G-scores, relative to the alternative of managers choosing players via ranking lists. They had managers choosing players based on ranking lists rather than randomly, and the G-score metric still worked well. 

Theoretically, one could explicitly model fantasy basketball as a game wherein managers choose their own ranking lists and draft players accordingly. It would be possible to determine exactly which players each manager would end up with, because the draft process would be entirely deterministic. This conception of the game obviates the need to assume that other players are chosen randomly. It is also a finite game, so by Nash's theorem it has a Nash Equilibrium (an equilibrium from which no manager has incentive to deviate). However, calculating the equilibrium would be computationally intractable. To calculate the Nash equilibrium, it would be necessary to know how each manager would fare under each possible strategy profile. Say that there were twelve teams with thirteen picks each, and there were $500$ total candidates. The number of possible rankings of relevant players would be

$$
\frac{500!}{ \left( 500 - 156 \right) !} \approx 10^{409}
$$

No modern computer is capable of this many operations in any amount of time. So while theoretically a Nash Equilibrium would exist, it would not be possible to actually calculate it 

\subsubsection{Performances sampled from distribution}

The set of possible performances $W$ represents a manager's belief about the probability distributions underlying future performance. $|W|$ can be infinite, making the probability distribution continuous, so any belief about player performance can be encoded with this model.

The only sense in which this distorts real fantasy basketball is that it does not account for players having their performance distributions change throughout a season. These changes can be significant, particularly if they are due to injury. Future work could consider their effects on value quantification

\subsubsection{No position requirement}

Most leagues have position requirements, which do not factor into the justification for G-scores. For fantasy basketball, the omission is not enormously problematic because there are often many flex spots, players are eligible for many positions, and value tends to be spread out fairly evenly between positions anyway. See Table \ref{fig:GDist} for the distribution from the 2022-2023 season.

\begin{table}[!ht]
\centering
\begin{tabular}{r r r r r r}
 & Center & Point Guard & Shooting Guard & Power Forward & Small Forward 	\\
 Rank & & & & & \\ \midrule
 1-52 & 13 & 12 & 10 & 9 & 8 \\
 53-104 & 10 & 11 & 13 & 11 & 6 \\
 105-156 & 7 & 10 & 11 & 15 & 8 \\
Overall & 30 & 34 & 34 & 36 & 22  \\
 \end{tabular}
\caption{Top 156 players by G-score in 2023}
\label{fig:GDist}
\end{table}

If a league has particularly tight position requirements, then a value-over-replacement (VOR) approach across positions may be warranted. VOR adjusts values by estimating how valuable the best player available after the draft would be for each position, then subtracts those values from all players of those positions. It is commonly used for fantasy football, which has tight position requirement (Fantasy Pros., 2017)

\subsubsection{Objective is number of categories won}

The expected value of number of categories won is not a perfect proxy for what managers want.

For one thing, it is obvious that this objective function is not correct for Most Categories. In that format, a manager would rather have a $100\%$ chance of winning five categories than a $99\%$ chance of winning all of them, even if they would win fewer categories on average. 

Another issue is that managers generally want to win their leagues or get close, which is not equivalent to maximizing the expected value of number of categories won. A manager with this preference would be inclined towards risky strategies that increase the probability of a strong performance while being sub-optimal for expected value. 

Neither of these issues is straightforward to incorporate into a value quantification model. With a static ranking list, there is no way of considering which categories a team is already weak or strong in, making it awkward to strategize around winning a majority of categories. And calculating the probability of winning a league would be much more complicated than calculating the probability of winning a single match-up.

Expected value of number of categories won is a reasonable compromise, because it gets at the main thrust of a managers' goal while facilitating a straightforward approach to modeling

\subsubsection{Player value is marginal relative to objective}

The idea of a value system is to use it to determine whether a player or group of players has more value than another. Ideally this calculation would consider player values in the context of each other, but that is impossible with a static system. Therefore, the only sensible way to aggregate value is to add marginal impacts of individual players

\subsection{General utility of static ranking lists}

Granting that the model described in Section \ref{Ass} represents the static value problem fairly, it should still be kept in mind that choosing players via a static ranking list is fundamentally sub-optimal. Drafting that way precludes the possibility of adapting based on draft circumstances, which is often advantageous.

One common adaptive strategy is called punting. When a manager punts, they strategically sacrifice one category in order to gain a consistent edge in others. To punt effectively, a manager needs to adapt to the players that happen to be available and form a punting strategy around them (Mamone, 2022). For example, if a player like Giannis Antetokounmpo who is generally strong but weak in free throw percentage is available, it suggests that punting free throws could be a profitable strategy. The same strategy would not work nearly as well if a player like Giannis was not available. It cannot be known beforehand whether Giannis will be available to draft, making it impossible to craft an optimal punting strategy before the draft has begun. Static ranking lists therefore cannot punt optimally.  

Another limitation of static ranking lists is that they can lead to unbalanced teams. For example, if the strongest players available to a drafter always happen to have extremely high rebound numbers, the drafter using a static ranking list would take all of them. This might allow them to dominate in rebounds but would not be optimal overall, since the team would be weak in other categories. Winning one category by a large margin and losing all others would result in a poor overall performance.

This is not to say that static value quantification is inherently useless. It can be useful for a context wherein managers are not capable of applying advanced strategies. Also, it can serve as a starting point for managers formulating more advanced strategies

\subsection{Consequences for head-to-head and Rotisserie}

Even assuming that long-term means are known perfectly, performances are volatile from from one week to another. This leads $\tau_M^2$ to be large for head to head formats, re-weighting G-scores in a significant way relative to Z-scores. Steals in particular are down-weighted heavily as demonstrated by Table \ref{fig:GCoef} based on real data from the 2022-2023 season.

\begin{table}[!ht]
    \begin{tabular}{r r r r r r r}
    & & & & & \\[-10pt]
    
        & $\sigma_M^2$ & $\tau_M^2$ & Z-score denominator & G-score denominator & G-score as fraction of Z-score 	\\ \midrule
        Assists & 41.57 & 31.55 & 6.45 & 8.55 & 75\% \\ 
        Blocks & 2.35 & 2.70 & 1.53 & 2.25 & 68\% \\ 
        Field Goal \% & 0.003 & 0.007 & 0.06 & 0.10 & 56\% \\ 
        Free Throw \% & 0.009 & 0.018 & 0.10 & 0.16 & 58\% \\ 
        Points & 325.52 & 448.32 & 18.04 & 27.82 & 65\% \\ 
        Rebounds & 52.01 & 57.55 & 7.21 & 10.47 & 69\% \\ 
        Steals & 1.01 & 4.20 & 1.00 & 2.28 & 44\% \\ 
        Threes & 9.52 & 9.04 & 3.09 & 4.31 & 72\% \\ 
        Turnovers & 5.45 & 8.85 & 2.34 & 3.78 & 62\% \\ 
    \end{tabular}
\caption{Empirical values for denominators of Z-score and G-score with real data from the 2022-2023 season, assuming random sampling from all weeks of the NBA season. $Q$ was calculated as the top 156 players by base Z-score, determined across the entire league}

\label{fig:GCoef}
\end{table}

Intuitively, this makes sense because volatile categories like steals are hard to win even with investment. Under the Z-score framework, an extra point of steals is not worth as much as an extra point of assists, because the value of the investment in steals is diluted by natural variance. G-scores account for this and make it so that scores for all categories have equivalent value.

Z-scores are not as problematic for Rotisserie, because weekly variations wash out to a degree across a full season. Still, using Z-scores for Rotisserie is not precisely optimal, since they ignore that some categories are harder to forecast for a full season than others

\subsection{Applying of the G-score formula}

\subsubsection{Defining Q}

There are a few sensible ways to define $Q$. One is to find a pseudo-equilibrium state in which $Q$ is composed of the top $|Q|$ players by G-score value. This can be accomplished by repeatedly re-calculating the top $|Q|$ players by G-score, setting $Q$ to them, and adjusting G-scores accordingly. If $Q$ does not change after an iteration, then it is an equilibrium set. 

It is also possible to estimate $Q$ using a popular ranking metric. This might be preferable in practice because G-scores are not necessarily reflective of how other managers actually value players 

\subsubsection{The $\kappa$ factor}

The $\kappa$ factor accounts for the fact that the player currently being chosen contributes additional period-to-period variance. This necessitates the inclusion of a small $\frac{2N}{2N-1}$ adjustment factor to period-to-period variance. 

Most leagues have twelve or thirteen players per team, so the $\kappa$ factor is generally between $\frac{26}{25} = 1.040$ and $\frac{24}{23} \approx 1.043$. As a multiplier it is therefore quite small, and can be held constant at $1.04$ or removed entirely for the sake of simplicity

\section{Conclusion}

G-scores, an extension of Z-scores, are justified via a set of simple assumptions. This improvement is especially important for head-to-head formats, because weekly variance can be significant, and ignoring it is heavily sub-optimal. 

There are many areas for future work, including optimizing picks dynamically based on drafting context. Static ranking lists are fundamentally sub-optimal relative to dynamic strategies so a truly sophisticated treatment of fantasy basketball would model some level of adaptation.

\textit{Disclaimer: The views and opinions expressed in this article are those of the independent author and do not represent those of any organization, company or entity}

\setcounter{secnumdepth}{-1}
\section{Appendix- Detailed justification of G-score}\label{apdx.G-score}

\subsection{Mathematical assumptions}

Several assumptions are invoked to simplify the justification behind G-scores. They are

\begin{itemize}
\item Category totals for a team can be estimated as Normal distributions due to the central limit theorem. Since there are usually $\approx 26$ relevant players, whose performances are within some standard range (fluctuations from 20 to 40 points per game are normal, but nobody is scoring 100), this is intuitively reasonable
\item The error function applied to a category total can be approximated with its first-order Taylor series
\item $\frac{X}{Y}$ can be approximated as a Normal distribution with $E \left( \frac{X}{Y} \right) \approx \frac{E(X)}{E(Y)}$. This assumes that $X$ and $Y$ are approximately Normal. independent, positive, and have low coefficient of variations (Diaz-Frances, 2012)
\item The variance of a ratio of success over attempts can be estimated as the first three terms in its Taylor series

\item  The period-to-period variance contribution of a player is approximately equal to the aggregate contribution

\begin{equation}
    \tau_M\left(p \right) \approx \tau_M
\label{eq4}
\end{equation} 

\begin{equation}
\tau_R \left( p \right) \approx \tau_R
\label{eq5}
\end{equation} 

\item Two quantities are approximately independent. Specifically, 

\begin{equation}
\frac{\sum_{ \left( q,w \right)  \in P} \theta_A \left( q, w \right)  }{{\sum_{ \left( q,w \right)  \in P} \left( \theta_A \left( q, w \right)  \right) + \theta_A \left(p,k \right)}} \perp \frac{\sum_{ \left( q,w \right)  \in P} \theta_A \left( q, w \right)  \theta_R \left( q, w \right)  - \sum_{q \in T_A} \theta_A \left( q, w \right)  \mu_R }{\sum_{ \left( q,w \right)  \in P} \theta_A \left( q, w \right) }
\label{eq3}
\end{equation}

\item The expected number of free throw attempts by a team where one player is known is approximately equal to the number of free throw attempts by an arbitrary team. Mathematically,

\begin{equation}
\mu_A * \left(N-1 \right) + \mu_A \left( p \right) \approx N \mu_A
\label{eq1}
\end{equation}

\item When calculating variance, the expected total number of free throw attempts by a team where one player is known is approximately equal to the number of free throws made by an arbitrary team. Mathematically,

\begin{equation}
\left( N-1 \right) \mu_A \mu_R + \mu_A \left( p \right) \mu_R\left( p \right) \approx N \mu_R \mu_A
\label{eq2}
\end{equation}

\end{itemize}

\subsection{Counting statistics}

A weekly total for a random player is essentially a random draw from a two-dimensional set of player and weekly performance. 

The mean of this draw is $\mu_M$. The variance can be calculated as 

$$\Var_{q,w} \left( \theta_M \left( q,w \right) \right)  =  \E_{q,w} \left(   \left( \theta_M \left( q,w \right) - \mu_M \right) ^2 \right)  =  \E_{q} \left(  \E_{w} \left(  \left( \theta_M \left( q,w \right) - \mu_M \right) ^2 \right)  \right)   
$$

Since $\E \left( X^2 \right)  = \Var \left( X \right)  + \E \left( X \right) ^2$, this translates to 

$$
\Var_{q,w} \left( \theta_M \left( q,w \right) \right) = \E_{q} \left(  \Var_{w} \left( \theta_M \left( q,w \right) - \mu_M \right)  + \left( \E_{w} \left( \theta_M \left( q,w \right) - \mu_M \right) \right) ^2 \right)   
$$

$$
= \E_{q} \left(  \Var_{w} \left( \theta_M \left( q,w \right) \right)  +  \left( \mu_M\left( q \right) - \mu_M \right) ^2 \right)
$$

$$
= \E_{q} \left( \tau_M \left( q \right)  \right)  + \Var_{q} \left( \mu_M\left( q \right) \right)  + \E_q \left(\mu_M\left( q \right) - \mu_M \right) ^2
$$

$$
= \E_{q} \left( \tau_M \left( q \right)  \right)  + \Var_{q} \left( \mu_M\left( q \right) \right)
$$

These two terms match the definitions of $\sigma_M^2$ and $\tau_M^2$ as defined by \ref{fig:GCountDef}, leading to a variance of

$$
\tau_M^2 + \sigma_M^2 
$$

Team $B$'s $N$ players in aggregate are then distributed with

\begin{itemize}
\item A mean of $N * \mu_M$
\item A variance of $ N *  \left( \sigma_M^2 + \tau_M^2 \right)$
\end{itemize}

Keeping in mind that the unchosen player's performance has mean $\mu_M\left( p \right)$ and variance $\tau_M\left(p \right)^2$ by definition, team $A$'s distribution has

\begin{itemize}
\item A mean of $\left(N-1 \right) * \mu_M + \mu_M\left( p \right)$
\item A variance of $\left(N-1 \right) *  \left( \sigma_M^2 + \tau_M^2 \right)  + \tau_M\left(p \right)^2$
\end{itemize}

Invoking the central limit theorem for team $B$'s total minus team $A$'s yields

$$
D \sim \mathcal{N} \left(  \mu_M - \mu_M\left( p \right),  \left( 2N-1 \right)  *  \left( \sigma_M^2 + \tau_M^2 \right)  + \tau_M\left(p \right)^2 \right) 
$$

The CDF of this quantity at zero is the probability of team $A$ winning the category. It is 

$$
   \frac{1}{2} \left[ 1 + \erf  \left(  \frac{ \mu_M\left( p \right) -  \mu_M}{\sqrt{2  \left( \left( 2N - 1 \right)  * \left( \sigma_M^2 + \tau_M^2 \right) + \tau_M \left( p \right)^2 \right)}}  \right)  \right]
$$

The first-order Taylor approximation, which is $\frac{2*X}{\sqrt{\pi}}$, can be used for the error function. This simplifies the expression to  

$$
   \frac{1}{2} \left[ 1 +\frac{2}{\sqrt{\pi}} *\frac{ \mu_M\left( p \right) - \mu_M}{ \sqrt{ 2 \left( (2N-1) (\sigma_M^2 + \tau_M^2) +  \tau_M\left(p \right)^2 \right) }} \right]
$$

Using the approximation of equation \ref{eq4} that $\tau_M\left(p \right) \approx \tau_M$ this becomes

$$
    \frac{1}{2} \left[ 1 +\frac{2}{\sqrt{\pi}} *\frac{ \mu_M\left( p \right) - \mu_M}{ \sqrt{ 2 \left( (2N-1) * \sigma_M^2 + 2N * \tau_M^2 \right) }} \right]
$$

$$
    \frac{1}{2} \left[ 1 +\frac{2}{\sqrt{\pi}} *\frac{ \mu_M\left( p \right) - \mu_M}{ \sqrt{ 2 \left( 2N- 1) \right) \left( \sigma_M^2 + \frac{2N}{2N - 1} * \tau_M^2 \right) }} \right]
$$

$$
   = \frac{1}{2} \left[ 1 +\frac{1}{\sqrt{\pi * \left(N - \frac{1}{2} \right)}  } *\frac{ \mu_M\left( p \right) - \mu_M}{ \sqrt{ \sigma_M^2 + \frac{2N}{2N-1} \tau_M^2  }} \right]
$$

\subsection{Percentage statistics}

The percentage statistics are more complicated to model
\subsubsection{Mean of the differential}

Consider $P$ to be a set of pairs $ \left(q,w \right) $ where $q$ is the chosen player and $w$ is the chosen week. Team $B$'s aggregate performance is a random draw from the two-dimensional distribution of player and scoring period  

$$
O_B = \frac{\sum_{ \left( q,w \right)  \in P} \theta_R \left( q, w \right)  \theta_A \left( q, w \right) }{\sum_{ \left( q,w \right)  \in P} \theta_A \left( q, w \right)   }
$$

Each element of the sum is equivalent. So 

$$
O_{\mu_B} = \frac{N * \E \left( \theta_R \left( q, w \right)  \theta_A \left( q, w \right)  \right) }{N * \E \left( \theta_A \left( q, w \right)  \right)  } =  \frac{ \E \left( \theta_R \left( q, w \right)  \theta_A \left( q, w \right)  \right) }{\E \left( \theta_A \left( q, w \right)  \right)  } = \frac{\E \left( \sum_{q \in Q, w \in W} \theta_R \left( q, w \right)  \theta_A \left( q, w \right)  \right) }{\E \left( \sum_{q \in Q, w \in W} \theta_A \left( q, w \right)  \right) }
$$

Since $\mu_A\left( q \right)$ is defined as $\frac{\sum_{w \in W} \theta_A \left( q, w \right) }{|W|}$, this can be rewritten to 

$$
O_{\mu_B} = \frac{ \E \left( \sum_{q \in Q,w \in W} \theta_A \left( q, w \right)  * \theta_R \left( q, w \right)  \right) }{ \E \left( |W| * \sum_{q \in Q} \mu_A\left( q \right) \right)}
$$

$\mu_R\left( q \right)$ is defined as $\frac{\sum_{w \in W} \theta_R \left( q, w \right)  * \theta_A \left( q, w \right) }{\sum_w \theta_A \left( q, w \right) }$ or 
$\frac{\sum_{w \in W} \theta_R \left( q, w \right)  * \theta_A \left( q, w \right) }{|W| * \mu_A\left( q \right)}$. That means $\sum_{w \in W} \theta_R \left( q, w \right)  * \theta_A \left( q, w \right)  = |W| * \mu_A\left( q \right) * \mu_R\left( q \right) $, allowing for a substitution to 

$$
O_{\mu_B} = \frac{ \E \left( \sum_{q \in Q} |W| * \mu_A\left( q \right) * \mu_R\left( q \right) \right) }{ \E \left( |W| * \sum_{q \in Q} \mu_A\left( q \right) \right)}  = \frac{ \E \left( \sum_{q \in Q} \mu_A\left( q \right) * \mu_R\left( q \right) \right) }{\E \left( \sum_{q \in Q} \mu_A\left( q \right) \right) }  = \mu_R
$$

Consider $k$ to be the randomly sampled week for the unchosen player $p$'s performance. Then, team A's success rate $O_A$ is 

$$
O_A = \frac{\sum_{ \left( q,w \right)  \in P} \theta_R \left( q, w \right)  \theta_A \left( q, w \right)  + \theta_R \left(k, p \right) \theta_A \left(p,k \right)}{\sum_{ \left( q,w \right)  \in P} \theta_A \left( q, w \right)   + \theta_A \left(p,k \right) }
$$

$O_A$ can be equivalently rewritten as

$$
\frac{\sum_{ \left( q,w \right) \in P} \theta_A \left( q, w \right)   \left( \theta_R \left( q, w \right)  - \mu_R \right)  + \sum_{ \left( q,w \right) \in P} \theta_A \left( q, w \right)  \mu_R + \theta_A \left(p,k \right)  \left( \theta_R \left(k, p \right) - \mu_R \right)  + \theta_A \left(p,k \right) \mu_R}{\sum_{ \left( q,w \right) \in P} \mu_A\left( q \right) + \theta_A \left(p,k \right)}
$$

Or broken into separate terms

\begin{align*}
O_A = &\frac{ \sum_{ \left( q,w \right) \in P} \theta_A \left( q, w \right)  \mu_R +  \theta_A \left(p,k \right) \mu_R}{{\sum_{ \left( q,w \right) \in P} \theta_A \left( q, w \right)  + \theta_A \left(p,k \right)}} + \frac{\sum_{ \left( q,w \right) \in P} \theta_A \left( q, w \right)   \left( \theta_R \left( q, w \right)  - \mu_R \right) }{{\sum_{ \left( q,w \right) \in P} \theta_A \left( q, w \right)  + \theta_A \left(p,k \right)}} \\
& + \frac{\theta_A \left(p,k \right)  \left( \theta_R \left(k, p \right) - \mu_R \right)  }{\sum_{ \left( q,w \right) \in P} \theta_A \left( q, w \right)  + \theta_A \left(p,k \right)}
\end{align*}

The first term can be simplified easily

\begin{align*}
& \mu_R \frac{ \sum_{ \left( q,w \right) \in P} \theta_A \left( q, w \right) +  \theta_A \left(p,k \right)}{{\sum_{ \left( q,w \right) \in P} \theta_A \left( q, w \right)  + \theta_A \left(p,k \right)}} + \frac{\sum_{ \left( q,w \right) \in P} \theta_A \left( q, w \right)   \left( \theta_R \left( q, w \right)  - \mu_R \right) }{{\sum_{ \left( q,w \right) \in P} \theta_A \left( q, w \right)  + \theta_A \left(p,k \right)}} 
\\ & + \frac{\theta_A \left(p,k \right)  \left( \theta_R \left(k, p \right) - \mu_R \right)  }{\sum_{ \left( q,w \right) \in P} \theta_A \left( q, w \right)  + \theta_A \left(p,k \right)} \\
= & \mu_R + \frac{\sum_{ \left( q,w \right) \in P} \theta_A \left( q, w \right)   \left( \theta_R \left( q, w \right)  - \mu_R \right) }{{\sum_{ \left( q,w \right) \in P} \theta_A \left( q, w \right)  + \theta_A \left(p,k \right)}}  + \frac{\theta_A \left(p,k \right)  \left( \theta_R \left(k, p \right) - \mu_R \right)  }{\sum_{ \left( q,w \right) \in P} \theta_A \left( q, w \right)  + \theta_A \left(p,k \right)}
\end{align*}

The expected value of team $A$'s success rate is then
$$
O_{\mu_A} = \mu_R + \E \left( - \frac{\sum_{ \left( q,w \right)  \in P} \theta_A \left( q, w \right)   \left( \theta_R \left( q, w \right) - \mu_R \right) }{{\sum_{ \left( q,w \right)  \in P} \theta_A \left( q, w \right)  + \theta_A \left(p,k \right)}} -  \frac{\theta_A \left(p,k \right)  \left( \theta_R \left(k, p \right) - \mu_R \right)  }{\sum_{ \left( q,w \right)  \in P} \theta_A \left( q, w \right)  + \theta_A \left(p,k \right)} \right) 
$$

To get the expected value of the differential, it is subtracted from $\mu_R$, leading to

\begin{align*}
D_\mu = & - \E \left( \frac{\sum_{ \left( q,w \right)  \in P} \theta_A \left( q, w \right)  \theta_R \left( q, w \right)  - \sum_{ \left( q,w \right)  \in P} \theta_A \left( q, w \right)  \mu_R  }{{\sum_{ \left( q,w \right)  \in P} \left( \theta_A \left( q, w \right)  \right) + \theta_A \left(p,k \right)}} \right) \\
& - \E \left( \frac{\theta_A \left(p,k \right)  \left( \theta_R \left(p,k \right) - \mu_R \right)  }{\sum_{ \left( q,w \right)  \in P} \left( \theta_A \left( q, w \right)  \right) + \theta_A \left(p,k \right)} \right)
\end{align*}

\begin{align*}
= &-\E \left( \frac{\sum_{ \left( q,w \right)  \in P} \theta_A \left( q, w \right)  }{{\sum_{ \left( q,w \right)  \in P} \left( \theta_A \left( q, w \right)  \right) + \theta_A \left(p,k \right)}} * \frac{\sum_{ \left( q,w \right)  \in P} \theta_A \left( q, w \right)  \theta_R \left( q, w \right)  - \sum_{ \left( q,w \right)  \in P} \theta_A \left( q, w \right)  \mu_R }{\sum_{ \left( q,w \right)  \in P} \theta_A \left( q, w \right) } \right)   \\
& -  \E \left( \frac{\theta_A \left(p,k \right)  \left( \theta_R \left(p,k \right) - \mu_R \right)  }{\sum_{ \left( q,w \right)  \in P} \left( \theta_A \left( q, w \right)  \right) + \theta_A \left(p,k \right)} \right)
\end{align*}

With the assumption of equation \ref{eq3}, this becomes 

\begin{align*}
D_\mu = & - \E \left( \frac{\sum_{ \left( q,w \right)  \in P} \theta_A \left( q, w \right)  }{{\sum_{ \left( q,w \right)  \in P} \left( \theta_A \left( q, w \right)  \right) + \theta_A \left(p,k \right)}} \right) \\
& \quad * \E \left( \frac{\sum_{ \left( q,w \right)  \in P} \theta_A \left( q, w \right)  \theta_R \left( q, w \right)  - \sum_{ \left( q,w \right)  \in P} \theta_A \left( q, w \right) \mu_R }{\sum_{ \left( q,w \right)  \in P} \theta_A \left( q, w \right) } \right)\\
& - \E \left[ \frac{\theta_A \left(p,k \right)  \left( \theta_R \left(p,k \right) - \mu_R \right)  }{\sum_{ \left( q,w \right)  \in P} \left( \theta_A \left( q, w \right)  \right) + \theta_A \left(p,k \right)} \right]
\end{align*}

The expected value of $\frac{\sum_{ \left( q,w \right) \in P} \theta_A \left( q, w \right)  \theta_R \left( q, w \right) }{\sum_{ \left( q,w \right) \in P} \theta_A \left( q, w \right) }$ is $\mu_R$ as proven for $O_B$, and $\frac{\sum_{q \in P} \theta_A \left( q, w \right)  \mu_R}{\sum_{ \left( q,w \right) \in P} \theta_A \left( q, w \right) }$ is $\mu_R$ because it can be pulled out of the sum as a constant. Therefore the first term cancels out and disappears. All that remains is the original second term,

$$
D_\mu = -  \E \left( \frac{\theta_A \left(p,k \right)  \left( \theta_R \left(k, p \right) - \mu_R \right)  }{\sum_{ \left( q,w \right) \in P} \theta_A \left( q, w \right)  + \theta_A \left(p,k \right)} \right) = -  \E \left( \frac{\theta_A \left(p,k \right) \theta_R \left(k, p \right) -\theta_A \left(p,k \right)  \mu_R }{\sum_{ \left( q,w \right) \in P} \theta_A \left( q, w \right)  + \theta_A \left(p,k \right)} \right) 
$$

Again applying that $E \left( \frac{X}{Y} \right) \approx \frac{E(X)}{E(Y)}$, the expression can be simplified to 

$$
D_\mu = - \frac{\mu_A\left( p \right)}{\left(N -1 \right) * \mu_A + \mu_A\left( p \right)}  \left( \mu_R\left( p \right) - \mu_R \right) 
$$

Further invoking the approximation of equation \ref{eq1}, 

$$
D_\mu = - \frac{\mu_A\left( p \right)}{N \mu_A}  \left( \mu_R\left( p \right) - \mu_R \right) 
$$

\subsubsection{Variance of the differential}

The variance requires the Taylor expansion, which is (Benaroya, 2005)

$$
\Var \left(  \frac{X}{Y} \right)  \approx \frac{\Var \left( X \right) }{\E \left( Y \right) ^2} - 2*\frac{\Cov \left( X,Y \right)  * \E \left( X \right) }{E \left( Y \right) ^3} + \frac{\E \left( X \right) ^2 * \Var \left( Y \right) }{\E \left( Y \right) ^4} 
$$

In the free throw case for team $B$, this becomes

\begin{align*}
O_{\sigma_B} ^2 = \frac{\Var \left(  \sum_{q \in T_B} \mu_A\left( q \right) \mu_R\left( q \right) \right) }{ \left( N \mu_A \right) ^2} & -  2 \frac{\Cov \left( \sum_{q \in T_B} \mu_A\left( q \right) \mu_R\left( q \right),\sum_{q \in T_B} \mu_A\left( q \right) \right)  * N \mu_A \mu_R }{ \left( N \mu_A \right) ^3} \\
& + \frac{ \left( N \mu_R \mu_A \right) ^2 * \Var \left( \sum_{q \in T_B} \mu_A\left( q \right)  \right) }{  \left( N \mu_A \right) ^4} 
\end{align*}

Or 

\begin{align*}
O_{\sigma_B} ^2 = \frac{\Var \left( \sum_{q \in T_B} \mu_A\left( q \right) \mu_R\left( q \right) \right) }{ \left( N \mu_A \right) ^2} & - 2 \frac{\Cov \left( \sum_{q \in T_B} \mu_A\left( q \right) \mu_R\left( q \right),\sum_{q \in T_B} \mu_A\left( q \right) \right)  \mu_R}{ \left( N \mu_A \right) ^2} \\
& + \frac{  \mu_R ^2 * \Var \left(  \sum_{q \in T_B} \mu_A\left( q \right) \right) }{  \left( N \mu_A \right) ^2} 
\end{align*}

Some basic facts about variance can be applied to the expression. $ C\Var \left( X \right)  = \Var \left( \sqrt{C}X \right) $,  and $C\Cov \left( X,Y \right)  = \Cov \left( \sqrt{C}X,\sqrt{C}Y \right)  = \Cov \left( X,CY \right) $, so the equation can be simplified to 

\begin{align*}
O_{\sigma_B} ^2 = &  \frac{\Var \left( \sum_{q \in T_B} \frac{\mu_A\left( q \right)}{\mu_A} \mu_R\left( q \right) \right)}{N^2} - \frac{2 \Cov \left( \sum_{q \in T_B} \frac{\mu_A\left( q \right)}{\mu_A} \mu_R\left( q \right),\sum_{q \in T_B} \frac{\mu_A\left( q \right)}{\mu_A} \mu_R \right)}{N^2} \\
& + \frac{\Var \left( \sum_{q \in T_B} \frac{\mu_A\left( q \right)}{\mu_A} \mu_R \right) }{N^2}
\end{align*}

It is known that $\Var \left( X \right)  + \Var \left( Y \right)  - 2 \Cov \left( X,Y \right) = \Var \left( X-Y \right) $ so this can be rewritten to

$$
O_{\sigma_B} ^2 = \frac{\Var \left( \sum_{q \in T_B} \frac{\mu_A\left( q \right)}{\mu_A} * \mu_R\left( q \right) - \frac{\mu_A\left( q \right)}{\mu_A} * \mu_R \right) }{N^2}
$$

Or 

$$
O_{\sigma_B} ^2 = \frac{\Var \left( \sum_{q \in T_B} \frac{\mu_A\left( q \right)}{\mu_A} *  \left( \mu_R\left( q \right) - \mu_R \right)  \right) }{N^2} = \frac{ \sum_{q \in T_B} \Var \left( \frac{\mu_A\left( q \right)}{\mu_A} *  \left( \mu_R\left( q \right) - \mu_R \right)  \right) }{N^2}
$$

There are $N$ players that contribute identical variance, so this can be rewritten to 

$$
O_{\sigma_B} ^2 = \frac{N * \Var_{q \in Q,w \in W} \left( \frac{\theta_A \left(q,w \right) }{\mu_A} *  \left( \theta_R \left( q, w \right)  - \mu_R \right)  \right) }{N^2}
$$

For the time being, define 

$$
\sigma_q^2 = \Var_{q \in Q,w \in W} \left( \frac{\theta_A \left(q,w \right) }{\mu_A} *  \left( \theta_R \left( q, w \right)  - \mu_R \right)  \right)
$$

$$
\sigma_p^2 = \Var_{w \in W} \left( \frac{\theta_A \left(p,w \right)}{\mu_A} *  \left( \theta_R \left(p,w \right) - \mu_R \right)  \right)
$$

The variance for team $B$ is then
$$
O_{\sigma_B} ^2 = \frac{N}{N^2} \sigma_q^2
= \frac{\sigma_q^2}{N}
$$ 

For team $A$, the variance is 

\begin{align*}
O_{\sigma_A} ^2 = & \frac{\Var \left(  \sum_{q \in T_A} \mu_A\left( q \right) \mu_R\left( q \right) \right) }{ \left( \left(N - 1 \right) \mu_A \right) ^2} \\
& - 2 \frac{\Cov \left( \sum_{q \in T_A} \mu_A\left( q \right) \mu_R\left( q \right),\sum_{q \in T_A} \mu_A\left( q \right) \right)  * \left( \left(N - 1 \right) \mu_A \mu_R + \mu_A\left( p \right) \mu_R\left( p \right) \right) }{ \left( N \mu_A \right) ^3} \\
& + \frac{ \left( \left(N - 1 \right) \mu_A \mu_R + \mu_A\left( p \right) \mu_R\left( p \right) \right) ^2 * \Var \left( \sum_{q \in T_A} \mu_A\left( q \right)  \right) }{  \left( N \mu_A \right) ^4} 
\end{align*}

Because of the assumption of equation \ref{eq2} that $ \left(N - 1 \right) \mu_A \mu_R + \mu_A\left( p \right) \mu_R\left( p \right) \approx N \mu_R \mu_A $, this becomes 

\begin{align*}
O_{\sigma_A} ^2 = & \frac{\Var \left(  \sum_{q \in T_A} \mu_A\left( q \right) \mu_R\left( q \right) \right) }{ \left( N \mu_A \right) ^2} -  2 \frac{\Cov \left( \sum_{q \in T_A} \mu_A\left( q \right) \mu_R\left( q \right),\sum_{q \in T_A} \mu_A\left( q \right) \right)  * \mu_R }{ \left( N \mu_A \right) ^2} \\
& + \frac{ \left( N \mu_A \mu_R \right) ^2 * \Var \left( \sum_{q \in T_A} \mu_A\left( q \right)  \right) }{  \left( N \mu_A \right) ^4} 
\end{align*}

Again some basic facts about variance can be applied to the expression. $ C\Var \left( X \right)  = \Var \left( \sqrt{C}X \right) $,  and $C\Cov \left( X,Y \right)  = \Cov \left( \sqrt{C}X,\sqrt{C}Y \right)  = \Cov \left( X,CY \right) $, so the equation can be simplified to 

\begin{align*}
O_{\sigma_A} ^2 = & \frac{\Var \left( \sum_{q \in T_A} \frac{\mu_A\left( q \right)}{\mu_A} \mu_R\left( q \right) \right)}{N^2} \\
& - \frac{2 \Cov \left( \sum_{q \in T_A} \frac{\mu_A\left( q \right)}{\mu_A} \mu_R\left( q \right),\sum_{q \in T_A} \frac{\mu_A\left( q \right)}{\mu_A} \mu_R \right)}{N^2}  + \frac{\Var \left( \sum_{q \in T_A} \frac{\mu_A\left( q \right)}{\mu_A} \mu_R \right) }{N^2}
\end{align*}

It is known that $\Var \left( X-Y \right)  = \Var \left( X \right)  + \Var \left( Y \right)  - 2 \Cov \left( X,Y \right) $ so this can be rewritten to

$$
O_{\sigma_A} ^2 = \frac{\Var \left( \sum_{q \in T_A} \frac{\mu_A\left( q \right)}{\mu_A} * \mu_R\left( q \right) - \frac{\mu_A\left( q \right)}{\mu_A} * \mu_R \right) }{N^2}
$$

In this case, there are $\left( N - 1 \right)$ players that contribute variance $\sigma_q^2$ and one that contributes $\sigma_p^2$, so this becomes 

$$
O_{\sigma_A} ^2 = \frac{\left(N - 1 \right) * \sigma_q^2 + \sigma_p^2 }{N^2}
$$

The variance of the differential is the variance of team A's score plus the variance of team B's score. 

$$
D_{\sigma}^2 = \frac{\left(N - 1 \right) * \sigma_q^2 + \sigma_p^2 }{N^2} + \frac{\sigma_q^2}{N}
= \frac{\left(2N - 1 \right) \sigma_q^2 + \sigma_p^2}{N^2 }
$$

Now $\sigma_q^2$ and $\sigma_p^2$ must be calculated. Starting with $\sigma_q^2$ , it is 

$$
\sigma_q^2=  \Var_{q \in Q,w \in W} \left( \frac{\theta_A \left(q,w \right) }{\mu_A}  \left( \theta_R \left( q, w \right)  - \mu_R \right)  \right) 
$$

$\sigma_q^2$ in this form can be easily computed if $W$ is available as a discrete set. It can also be translated to a more understandable form.

Since $\Var \left( X \right)  = \E \left( X^2 \right)  + \E \left( X \right) ^2$ and the expected value of $\frac{\theta_A \left(q,w \right) }{\mu_A}  \left( \theta_R \left( q, w \right)  - \mu_R \right)  =0$, this becomes 

$$
\sigma_q^2=  \E_{q \in Q, w \in W} \left( \frac{\theta_A \left(q,w \right) ^2}{\mu_A^2}  \left( \theta_R \left( q, w \right)  - \mu_R \right) ^2 \right) 
$$

$$
= \E_q\left( \E_w \left( \frac{\theta_A \left(q,w \right) ^2}{\mu_A^2}  \left( \theta_R \left( q, w \right)  - \mu_R \right) ^2 \right) \right) 
$$

$\E \left( X^2 \right)  = \Var \left( X \right)  - \E \left( X \right) ^2$ so this can be translated into 

$$
\sigma_q^2 =  \E_q \left( \Var_w \left( \frac{\theta_A \left(q,w \right) }{\mu_A}  \left( \theta_R \left( q, w \right)  - \mu_R \right)  \right) + \E_w \left( \frac{\theta_A \left(q,w \right) }{\mu_A}  \left( \theta_R \left( q, w \right)  - \mu_R \right)  \right) ^2 \right) 
$$

$$
= \ \E_q \left( \Var_w \left( \frac{\theta_A \left(q,w \right) }{\mu_A}  \left( \theta_R \left( q, w \right)  - \mu_R \right)  \right)+ \E_w  \left( \frac{\theta_A \left(q,w \right)  \theta_R \left( q, w \right) }{\mu_A} - \frac{\theta_A \left(q,w \right)  \mu_R}{\mu_A} \right)  ^2 \right) 
$$ 

$$
= \E_q \left( \Var_w \left( \frac{\theta_A \left(q,w \right) }{\mu_A}  \left( \theta_R \left( q, w \right)  - \mu_R \right) \right) + \left( \E_w  \left( \frac{\theta_A \left(q,w \right)  \theta_R \left( q, w \right) }{\mu_A} \right)  - \E_w \left( \frac{\theta_A \left(q,w \right)  \mu_R}{\mu_A} \right)  \right)^2  \right)
$$ 

$$
= \E_q \left(  \Var_w \left( \frac{\theta_A \left(q,w \right) }{\mu_A}  \left( \theta_R \left( q, w \right)  - \mu_R \right) \right) +  \left( \frac{\mu_A\left( q \right) \mu_R\left( q \right)}{\mu_A} - \frac{\mu_A\left( q \right) \mu_R}{\mu_A} \right)^2 \right) 
$$ 

$$
=  \E_q \left( \Var_w \left( \frac{\theta_A \left(q,w \right) }{\mu_A}  \left( \theta_R \left( q, w \right)  - \mu_R \right) \right) +  \left( \frac{\mu_A\left( q \right)}{\mu_A}  \left( \mu_R\left( q \right) - \mu_R \right)  \right) ^2  \right) 
$$ 

Separating out the two terms yields 

\begin{align*}
\sigma_q^2 =  & \left( \E_q \Var_w \left( \frac{\theta_A \left(q,w \right) }{\mu_A}  \left( \theta_R \left( q, w \right)  - \mu_R \right)  \right) + \E_q  \left( \frac{\mu_A\left( q \right)}{\mu_A}  \left( \mu_R\left( q \right) - \mu_R \right)  \right) ^2 \right) \\
& = \E_q \left( \Var_w \left( \frac{\theta_A \left(q,w \right) }{\mu_A}  \left( \theta_R \left( q, w \right)  - \mu_R \right) \right) \right){N} + \Var_q   \left( \frac{\mu_A\left( q \right)}{\mu_A}  \left( \mu_R\left( q \right) - \mu_R \right)  \right) \\
& \quad + \left( \E \left( \frac{\mu_A\left( q \right)}{\mu_A}  \left( \mu_R\left( q \right) - \mu_R \right)  \right) \right) ^2 
\end{align*}

Note that $\E \left( \frac{\mu_A\left( q \right)}{\mu_A}  \left( \mu_R\left( q \right) - \mu_R \right)  \right)  = \E \left( \frac{\mu_A\left( q \right) * \mu_R\left( q \right)}{\mu_A} \right)  - \E \left( \frac{\mu_A\left( q \right) * \mu_R}{\mu_A} \right)  = \mu_R - \frac{\mu_A * \mu_R}{\mu_A} = \mu_R - \mu_R = 0$. That leads to a form of 

$$
\sigma_q^2 =  \left( \E_q \left( \Var_w \left( \frac{\theta_A \left(q,w \right) }{\mu_A}  \left( \theta_R \left( q, w \right)  - \mu_R \right) \right) \right)  + \Var_q \left(  \left( \frac{\mu_A\left( q \right)}{\mu_A}  \left( \mu_R\left( q \right) - \mu_R \right)  \right)  \right)  \right) 
$$ 

The first term matches $\tau_R^2$ and the second term matches $\sigma_R^2$. The resulting overall variance is then 

$$
\sigma_q^2 = \sigma_R^2 + \tau_R^2
$$

For $\sigma_p^2$, because of equation \ref{eq5}

$$
\sigma_p^2 =  \tau_R \left(p \right) ^2 \approx \tau_R^2
$$

And finally 

$$
D_{\sigma}^2 = 
= \frac{\left(2N - 1 \right) \sigma_q^2 + \sigma_p^2}{N^2 }
=  \frac{\left(2N - 1 \right) \left( \sigma_R^2 + \tau_R^2 \right) + \tau_R^2}{N^2 }
= \frac{\left(2N - 1 \right)  \sigma_R^2 + 2N \tau_R^2}{N^2 }
$$

\subsubsection{Differential distribution}

Now the normal approximation CDF at zero can be written
$$
   \frac{1}{2} \left[ 1 + \erf \left( \frac{ \frac{\mu_A\left( p \right)}{N \mu_A}  \left( \mu_R\left( p \right) - \mu_R \right) }{\sqrt{2 * \frac{\left(2N - 1 \right)  \sigma_R^2 + 2N \tau_R^2}{N^2 }  }} \right) \right]
$$

Applying the Taylor first-order approximation yields 

$$
   \frac{1}{2} \left[ 1 + \frac{2}{\sqrt{\pi}} * \frac{ \frac{\mu_A\left( p \right)}{N \mu_A}  \left( \mu_R\left( p \right) - \mu_R \right) }{\sqrt{2 * \frac{\left(2N - 1 \right)  \sigma_R^2 + 2N \tau_R^2}{N^2 } }} \right]
$$

$$
   = \frac{1}{2} \left[ 1 + \frac{2}{\sqrt{\pi}} * \frac{ \frac{\mu_A\left( p \right)}{\mu_A N }  \left( \mu_R\left( p \right) - \mu_R \right) }{\sqrt{2 * \frac{\left(2N - 1 \right)  \sigma_R^2 + 2N \tau_R^2}{N^2 } }} \right]
$$

$$
   = \frac{1}{2} \left[ 1 + \frac{1}{\sqrt{\pi \left(N - \frac{1}{2} \right) }} * \frac{ \frac{\mu_A\left( p \right)}{\mu_A}  \left( \mu_R\left( p \right) - \mu_R \right) }{\sqrt{  \sigma_R^2 + \frac{2N}{2N-1} \tau_R^2 }} \right]
$$

\subsection{Total across categories}

Adding together the G-scores for all counting statistics percentage statistics yields 

$$
   \frac{1}{2}\left[|C| + \frac{1}{\sqrt{\pi * \left( N - \frac{1}{2} \right) }} * \sum_c G_c \right]
$$

Where $G_c$ is the G-score for a category.

This is player value, as defined in Section \ref{Model}.  G-score is clearly proportional to it, with each point of G-score being worth 

$$
    \frac{1}{2 \sqrt{\pi * \left( N - \frac{1}{2} \right)}}
$$

So G-score meets the criteria for player value as well

\bibliographystyle{agsm}

\end{document}